\documentstyle[12pt,aaspp4]{article} 

\lefthead{Nakasato et al.}
\righthead{GRAPE-SPH with Parallel Virtual Machine}

\begin{document}

\title{Smoothed Particle Hydrodynamics with GRAPE
and Parallel Virtual Machine}

\author{Naohito Nakasato\altaffilmark{1},Masao Mori\altaffilmark{2},
and Ken'ichi Nomoto\altaffilmark{1,3}}

\altaffiltext{1}{Department of Astronomy, School of Science,
University of Tokyo, Tokyo 113}

\altaffiltext{2}{The Institute of Astronomy, Faculty of Science, 
University of Tokyo, Mitaka, Tokyo 181, Japan}

\altaffiltext{3}{Research Center for the Early Universe,
School of Science, University of Tokyo, Tokyo 113}

\begin{abstract}
We have developed Remote-GRAPE, a subroutine library
to use the special purpose computer GRAPE-3A.
The GRAPE-3A can efficiently calculate gravitational
force between particles, and construct neighbor lists.
All other calculations are performed on the host workstation (WS) which
is directly connected to GRAPE. The use of GRAPE for Smoothed Particle
Hydrodynamics (GRAPE-SPH) can in principle greatly speed up the
calculations on WS.
However the current bottleneck of GRAPE-SPH is that
its performance is limited by the speed of the host WS.
To solve this problem, we implement Remote-GRAPE;
it allows us to run applications which use GRAPE-3A hardware
on the significantly faster computers than the physical host WS.
Thus, we can take advantage of the fast computers even though
they can not physically be connected to GRAPE.
The Remote-GRAPE system is implemented on the Parallel Virtual Machine (PVM).
The detail of implementation is described.

We analyze the performance of Remote-GRAPE and obtain the following results.
1) When Remote-GRAPE is used to calculate only gravitational forces,
the overhead due to the network
decrease according to the number of particles (20 \% - 40 \% of total time).
2) To calculate gravity and neighbor lists,
the overhead due to the network does not occupy a large fraction of total time
but only $\sim$ 20 - 30 \%, because the computation required
on the slave machine (see text) is very large due to the property of
GRAPE system.
3) We also compare the performance of Remote-GRAPE with the tree method.
The tree method requires more time for higher degree of clustering, 
while the required time for Remote-GRAPE does not much depend on it.

We then analyze the performance of GRAPE-SPH with Remote-GRAPE.
Its performance is about 4 times faster than GRAPE-SPH with usual GRAPE
for our configuration.
Using Remote-GRAPE, we can calculate the GRAPE part and other parts in 
parallel. This parallel method leads to further speed up of our SPH code.
We estimate how the performance of Remote-GRAPE depends on the configuration.
We also show that the performance of Remote-GRAPE can be further improved.
\end{abstract}


\clearpage

\section{Introduction}

The Smoothed Particle Hydrodynamics method (SPH) is widely used
to calculate three dimensional hydrodynamics with Lagrange scheme
(Lucy 1977; Gingold \& Monaghan 1977).
It has been applied to many astrophysical problems.
Because of its Lagrangian nature, it is suitable to the problem
which has large density contrasts,
e.g. the formation of galaxies(Evrard 1988; Hernquist \& Katz 1989;
Umemura 1993; Steinmetz \& Mueller 1994) or a cloud-cloud collision
(Lattanzio et al. 1985; Habe \& Ohta 1992).
In these calculation, in order to treat star particles and/or
dark matter with gas particle, we have to solve SPH with
collision-less particles (N-body system).
Various codes have been developed to combine SPH and N-body system.
In these codes, gravitational forces are calculated
in many different ways such as direct summations,
Particle-Particle/Particle-Mesh methods (Evrard 1988), Tree methods
(Hernquist \& Katz 1989; Benz et al. 1990),
and the method to use the special purpose computer GRAPE
(Umemura et al. 1993; Steinmetz 1996).

GRAPE (GRAvity PipE) is a special purpose computer for efficiently
calculating gravitational force and potential (Sugimoto et al. 1990).
We need a host computer, which is connected to GRAPE board, to
control it and conduct other calculations (e.g. time integration).
The use of GRAPE for SPH simulations (hereafter GRAPE-SPH) has many
advantages, since GRAPE can calculate not only
gravitational force and potential but also construct
lists of neighbor particles in a short time.
In SPH simulations we need the lists of neighbor particles to calculate
hydrodynamical quantities. Searching neighbor particles with GRAPE is
much more efficient than a direct search on the host computer.
With GRAPE-SPH, therefore, we need to calculate only pure hydrodynamical
part of SPH on the host computer. 

Although the cost of N-body part of SPH can be significantly reduced
by using GRAPE, the speed of the hydrodynamical part of the code
is limited by the speed of the host computer.
The speed of workstation (WS) is being rapidly improved.
To take full advantage of GRAPE for SPH calculations, 
we have to use the state-of-the-art WS as a host.
If the host WS of GRAPE is much slower than a fast WS available now, 
the total performance of GRAPE-SPH is lower than that of the SPH simulation
on the fast WS without GRAPE.
Of course, by developing new interface to new WS, we could solve this problem.
This approach, however, requires many human time.
Moreover, during the development of the new interface,an ever newer WS
with different interface might become available.

\begin{figure}[t]
\begin{center}\leavevmode
\epsfxsize=0.8\columnwidth
\epsfbox{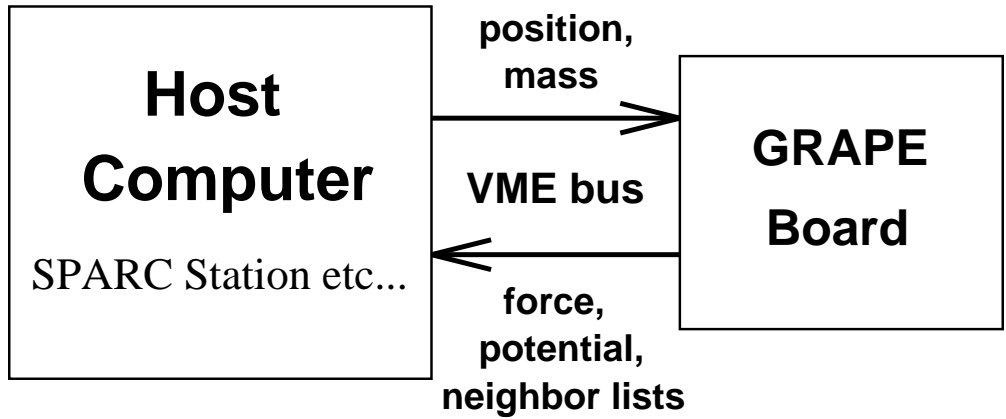}
\end{center}
\caption{
The schematic diagram of the GRAPE system.\label{fig1}
}
\end{figure}

\begin{figure}
\begin{center}\leavevmode
\epsfxsize=0.8\columnwidth
\epsfbox{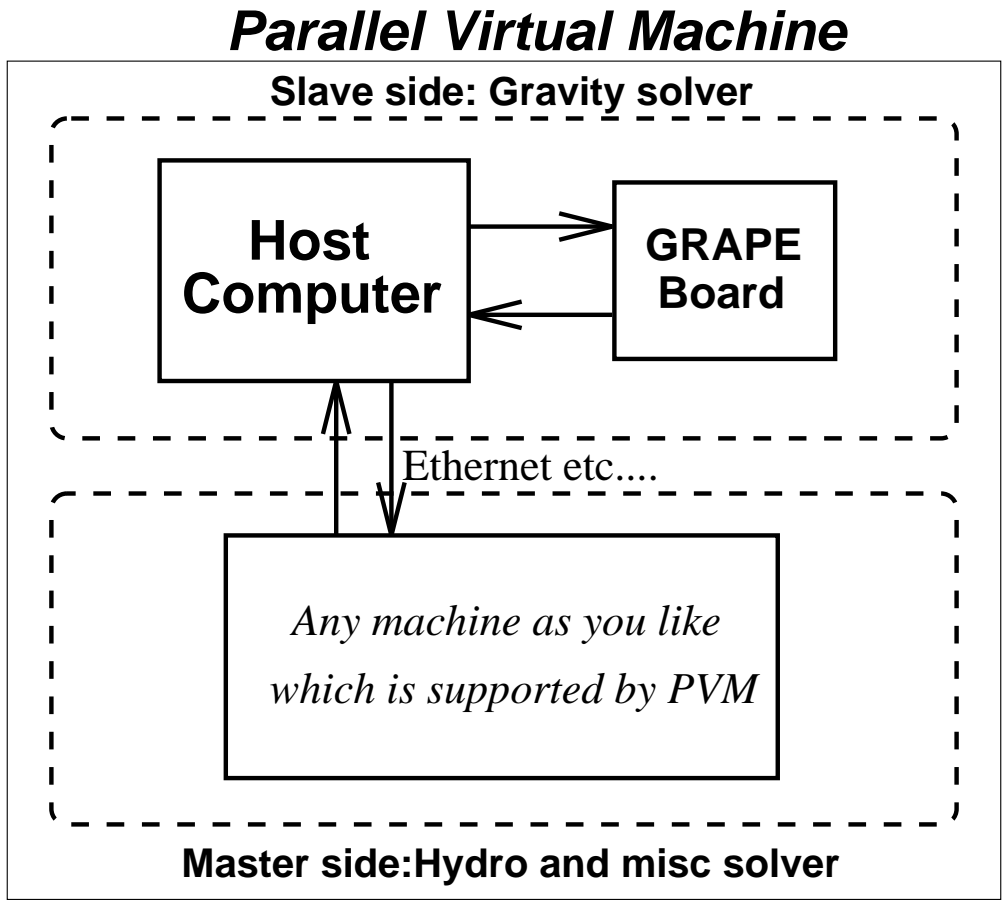}
\end{center}
\caption{
The concept of PVM and our library.\label{fig2}
}
\end{figure}

Here we take a different, novel approach to solve this problem.
We use Parallel Virtual Machine (PVM; Geist et al. 1994),
which is one of the most
popular message-passing systems in parallel computing, 
to connect the WS which is directly connected to a GRAPE board
with another fast machine. The SPH part is performed on the fast machine. 
In fact, all simulation code is run on the fast machine
and the WS connected to GRAPE board serves essentially as the intelligent
communication interface.
Figure \ref{fig1} and \ref{fig2} shows the present GRAPE system and
our new approach, respectively.
In our new approach, the combination of the WS and GRAPE
(Figure \ref{fig1}) behave as the ``remote GRAPE'' system
which is connected directly to the Local Area Network (LAN).
Thus, any computer on the LAN can be used as the host
(remote-host) of the GRAPE.
We construct the library named Remote-GRAPE.
In this paper we present the implementation of Remote-GRAPE
and discuss its performance.
In section 2 we summarize the GRAPE system and the bottleneck of GRAPE-SPH.
In section 3 we describe how the Remote-GRAPE is implemented, analyze
its performance, and compare it with the original GRAPE system
and other scheme.
In section 4 we show the performance of our GRAPE-SPH code and
discuss possible improvement of our code.
In section 5 conclusions are summarized.

\begin{figure}[t]
\begin{center}\leavevmode
\epsfxsize=0.8\columnwidth
\epsfbox{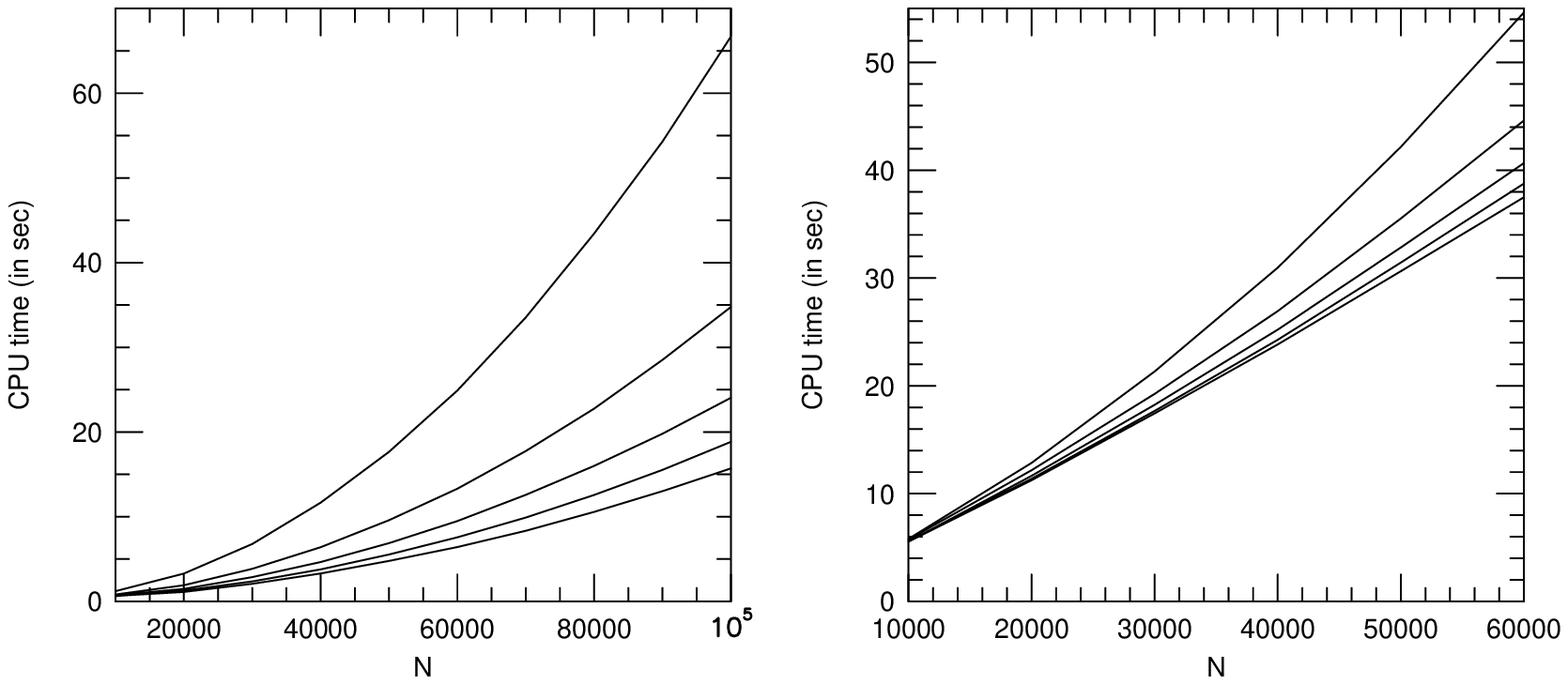}
\end{center}
\caption{
(a) left : The CPU time as a function of the number of particles $N$
in calculating only gravitational force (Case G).
The five solid lines correspond to the number of GRAPE boards of 1 to 5
from top to bottom.
(b) right : Same as Figure 3a, but for calculating both gravitational
force and neighbor lists (Case GN).
In both cases the GRAPE boards are connected to SPARC Classic.
\label{grape1}
}
\end{figure}

\section{The GRAPE and GRAPE-SPH}
The GRAPE-3AF system (Okumura et al. 1993) has 8 processors on one board, 
and the GRAPE board is connected to a host computer via VME
(Versa Module Europe) bus.
The processor chip is designed to calculate gravitational
force with a Plummer softening, i.e., 
\begin{equation} 
\mbox{\boldmath{$f$}}( \mbox{\boldmath{$r$}}_i ) =
-G \sum_{j}^{N} m_j
\frac{ \mbox{\boldmath{$r$}}_i - \mbox{\boldmath{$r$}}_j}
{(\mbox{\boldmath{$r$}}_{ij}^2 +\epsilon^2)^{1.5}}.
\end{equation}
During the force calculation, GRAPE-3AF can construct
neighbor lists simultaneously (Fukushige et al. 1991; Okumura et al. 1993).
The size of the buffer for neighbor lists is 1024$\times$4byte, 
which limits the maximum number of neighbors per board.
The schematic diagram of the GRAPE system is shown in Figure \ref{fig1}.

We present general performance of GRAPE system
in Figures \ref{grape1}a and \ref{grape1}b.
These figures show the relation between the CPU time and number of
particles, $N$, for different number of GRAPE boards.
In our site, the GRAPE boards are connected to Sun SPARC Classic
(hereafter Classic) via Aval Data SVA-100 VME interface.
Figure \ref{grape1}a (left) shows the case that calculates
only gravity (hereafter Case G), while Figure \ref{grape1}b (right) shows
the case that calculates both gravity and neighbor lists (Case GN).
The particles used in these experiments are randomly distributed
within the calculation region (see Figure \ref{data}a), 
and the size of the region changes with $N$ to make the number density
almost constant.
The average number of neighbor particles is almost constant
($\sim$ 40) for any $N$.

\begin{figure}[t]
\begin{center}\leavevmode
\epsfxsize=0.8\columnwidth
\epsfbox{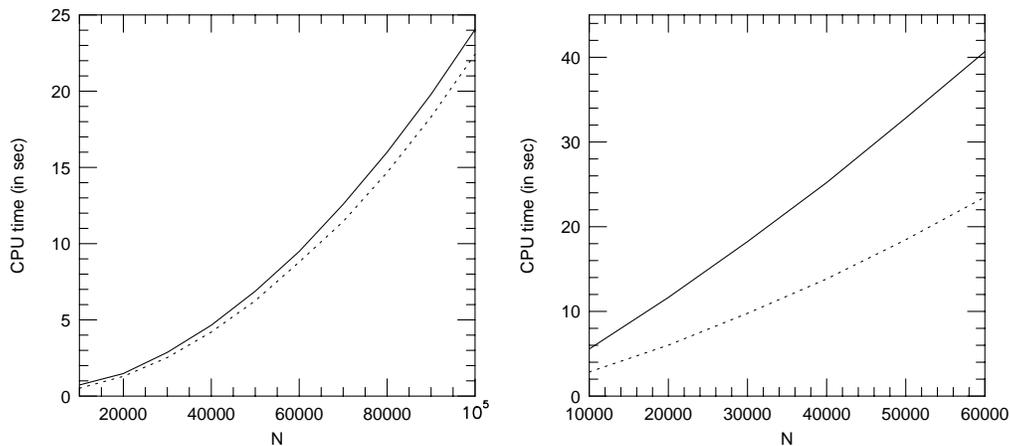}
\end{center}
\caption{
The CPU time as a function of $N$ where SPARC Classic (solid) and
SPARC Station 10 (dotted) are used as a host computer of GRAPE.
The number of GRAPE board is three for both machine.
(a) left : Case G. (b) right : Case GN.
\label{grape2}
}
\end{figure}

In Case G, the performance is dramatically improved by
increasing the number of boards.
In Case GN, however, the use of larger number of boards does not
significantly improve the performance.
This is because the speed of the VME bus is too slow
to read neighbor lists for large $N$ with many boards, and
the calculations required to construct neighbor lists
on the host computer is rather costly work for our machine.
We compare the performances of GRAPE system with different host computers. 
We use Classic (cf. SPEC int92/fp92 : 26.4/21.0,
which represent integer and floating performance of the machine)
and Sun SPARC Station 10/40 (hereafter SS-10 with int92/fp92 : 50.2/60.2)
with the same interface and three GRAPE boards.
Figure \ref{grape2}a and \ref{grape2}b show cases G and GN, respectively,
where the solid and dotted lines represent the cases of
Classic and SS-10, respectively.
For Case GN, the computation on host computer
(in short, construction of neighbor lists) is longer than Case G.
Thus the speed of the host computer apparently affect the performance.

In SPH, physical quantities at one position are calculated by
smoothly averaging over neighbor particles. 
Thus, SPH simulations are essentially equivalent to
N-body simulation with complex, short range ``force''.
In GRAPE-SPH we use GRAPE for calculating gravitational force and searching
neighbor particles.
Steinmetz (1996) summarized the performance of GRAPE-SPH in his table 1
for one GRAPE-3AF board and Sun SPARC Station 10 as a host computer.
He showed that about 80 \% ($N \sim 60000$) of computing time is spent on
the hydrodynamical and miscellaneous calculations, 
and only $\sim$ 20 \% is spent on the GRAPE part.
This implies that a higher performance is obtained
if we can use a faster WS to calculate the SPH part.

\begin{table*}[t]
\caption{Description of test cases for GRAPE and Remote-GRAPE.
\label{table_grape}
}
\begin{center}
\begin{tabular}{ccccc}
\hline\hline
Case & Gravity$^{\rm a}$ & Neighbor lists$^{\rm a}$ &
Remote$^{\rm b}$ & Section$^{\rm c}$ \\
\hline
G   & $\bigcirc$ & $\times$    & $\times$   & 2  \\
GN  & $\bigcirc$ & $\bigcirc$  & $\times$   & 2  \\
RG  & $\bigcirc$ & $\times$    & $\bigcirc$ & 3.3.1\\
RGN & $\bigcirc$ & $\bigcirc$  & $\bigcirc$ & 3.3.2\\
\hline
\end{tabular}
\\
$^{\rm a}$These column indicate whether neighbor lists (N) are calculated
in addition to gravity (G).\\
$^{\rm b}$This column indicates whether Remote-GRAPE is used.\\
$^{\rm c}$This column indicates which section the test case is discussed.\\
\end{center}
\end{table*}

\section{Implementation of Remote-GRAPE and its Performance}

\subsection{The motivation to develop the Remote-GRAPE}
The current bottleneck of the performance of GRAPE-SPH is
the slow host computer, while the choice of the host computer
is limited by the interface of GRAPE-3AF.
Our approach to solve this problem is to develop the software which
allows us to use any computer connected to the network as a host of GRAPE.
This makes it possible to use a host computer of
any vendors and of any performance. For this purpose,
we adopt the Parallel Virtual Machine as a basis of the code.

The Parallel Virtual Machine (PVM) is a software that enable us to develop
a highly portable parallel program with the message-passing style
(Geist et al. 1994). We can solve large computational problems by
using a heterogeneous collection of computers as a single parallel computer.
In PVM, subroutines to send and receive messages can be called from
C language or Fortran. Here the message means that the data which
are needed for solving the problem.
We name the library "Remote-GRAPE".

\subsection{Implementation of Remote-GRAPE}

We call a machine on which the main program runs as a {\tt master}
machine or {\tt master} side, and a machine which is connected physically
to GRAPE boards as a {\tt slave} machine or {\tt slave} side.
Also we call the program which runs on the master machine
as {\sl master} task and the program which runs on the slave machine
as {\sl slave} task. Figure \ref{fig2} shows the concept of our method.

In the library of GRAPE system, a simple example code looks like 
the following (see Makino \& Funato 1993 for details):
\begin{verbatim}
real scale_leng, scale_mass, eps2
real real x(3,nall), mass(nall)
real acc(3,nall), pot(nall)
.........
call g3setscales(scale_leng, scale_mass)
call g3seteps2(eps2)
call g3setn(nall)
do j = 1, nall
call g3setxj(j-1,x(1,j))
call g3setm(j-1,mass(j))
enddo
do i = 1, nall, 8
call g3frc(x(1,i), acc(1,i), pot(i), 8)
enddo
\end{verbatim}
In the first three lines we set the scales of length and mass,
the softening parameter, and the number of particles, respectively.
Then we set positions and mass of the particles in the first {\tt do} loop and
call {\tt g3frc()} to get gravitational force and potential in 
the second {\tt do} loop.

When we are sending a message in PVM, we may add an identifier to the message.
This makes it possible for the program, which is receiving the message,
to distinguish one message from another.
We adopt this mechanism at the slave task to select the routine
which the master task wants to call.
For example, if the master task wants to call {\tt g3setn()}, 
it has to send the message, which contains the number of particles, 
to the slave task with previously defined tag for {\tt g3setn()}.
The slave task will receive the message from the master task
and examine the tag, and then call suitable routine, 
{\tt g3setn()} in this case.

With this method we can call essentially any routine in the original
GRAPE library from any machine.
If small messages are sent many times, however, 
it requires a large computational and network cost.
For example, it is more efficient to make a new routine
which gathers many calls than to call g3setxj() separately for each particle.
Thus in our library, we send and receive the data
of many particles in one message.
Moreover we compile calling of many routines like
the second {\tt do} loop in one routine.
A simple example of the code using Remote-GRAPE is as follows:
\begin{verbatim}
real scale_leng, scale_mass, eps2
real x(3,nall), mass(nall), acc(3,nall),
real pot(nall), h(nall) 
integer*2 list(nall*100)
integer   begin(nall)
.........
call g3setscales(scale_leng, scale_mass)
call g3seteps2(eps2)
call g3setmass(nall,mass)
call g3setdata(nall,x(1,1), x(2,1), x(3,1))
call g3seth2all(nall, h)
call g3docalculation(1,nall,1, nall, 2)
call g3getgravity(nall, acc(1,1), acc(2,1),
     acc(3,1), pot)
call g3getneighbor(nall, list, begin)
\end{verbatim}
The first two routines are the same as the original GRAPE library.
The next three routines, {\tt g3setmass()}, {\tt g3setdata()},
and {\tt g3seth2all} set, respectively, the mass, position,
and the neighbor searching radius of all particles.
These routines only send the data, and the slave task stores the data in 
internal working space.
Then calling {\tt g3docalculation()} starts calculations.
At this time, the slave task perform the codes 
which is similar to the original example.
Finally, we get gravitational force, potential energy, and neighbor lists.

The arrays {\tt list} and {\tt begin} are neighbor lists.
The array {\tt list} is whole list of neighbor particles,
and array {\tt begin} is a list of the index which represents 
the first member of a neighbor list of a particle.
For example if the first part of the {\tt begin} looks like
(1, 24, 47, 58, ...), elements 1 to 23 of the array {\tt list} are
the neighbor particles of particle 1, 
elements 24 to 46 are the neighbors of particle 2, and so on. 
When overflow of neighbor lists occurs in the slave task routine
of constructing neighbor lists, we use the same method as adopted
by Steinmetz (1996).

Since the throughput of the message transfer is rather low,
we minimize the amount of data to be transferred.
Though the original GRAPE library has both single precision and
double precision routines, the accuracy of GRAPE itself is limited
approximately to the single precision.
Thus, to get the best performance of Remote-GRAPE,
we use all messages in our library in single precision.
In case we need double precision, we set a double to single precision
conversion factor, 
because the data transfer is always done in single precision.
A required cost to convert single to double is small.
For the same reason we use {\sl unsigned short} (2byte) for neighbor lists.
When neighbor lists are used, the number of particles 
is limited to $\sim$ 64,000.

\subsection{The performance of Remote-GRAPE}
Here we analyze the performance of Remote-GRAPE.
The benefit of using Remote-GRAPE over usual GRAPE
is that Remote-GRAPE is almost machine independent
as far as the machine is supported by PVM.
In compensation, we have to transfer the data
across the LAN, which is usually slower than the bus that connects 
GRAPE board and the host computer.
Here we compare the performance of Remote-GRAPE and usual GRAPE.

In our site, the master machine is DEC Alpha station 600 5/266
(cf. int92/fp92 : 292.8/433.5), 
and the slave machine is Sun SPARC Classic with five GRAPE-3AF boards. 
Hereafter all values are obtained from timing on these machines,
and the timing is not CPU time but elapsed time.
The reason why we present elapse time is that,
if we use usual routines or commands
to get CPU time in the program using Remote-GRAPE library, 
it only returns CPU time on master side. 
This time does not include the time to spend on slave side and
to transfer data over the LAN, so it is meaningless for our purpose.
When we examine the performance, there are a few or none active process
except our interested program. Also we use both the Ethernet (10Mbps) and 
the Fast-Ethernet (100Mbps) as the backbone of LAN.
The difference between the two backbone is also analyzed.
We summarize test cases for GRAPE and Remote-GRAPE in Table
\ref{table_grape}.

\subsubsection{Case RG : Only gravity}
For $N$ particles, the time require to calculate gravitational forces
by GRAPE is expressed as
\begin{equation}
t_{\rm G} = a_1 N^2 + (a_2 + a_3) N,
\end{equation}
where $a_1$ is the time required to calculate gravity of one particle, 
$a_2$ the time to transfer data via VME bus, and
$a_3$ the time spent for miscellaneous calculations on the slave machine.
The time require to transfer messages via the LAN is expressed as
\begin{equation}
t_{\rm LAN} = a_4 N,
\end{equation}
where $a_4$ is the time for one particle.
Using Remote-GRAPE, the total required time is given as
\begin{equation}
t_{\rm RG} = t_{\rm G} + t_{\rm LAN}.
\end{equation}
We estimate the values of $a_1, a_2$ and $a_3$ as
\begin{eqnarray}
&a_1 = 5.0 \times 10^{-8}/N_{chip} \sim 1.3 \times 10^{-9} \ {\rm sec},
&\nonumber\\
&a_2 \sim 2.0 \times 10^{-5} \ {\rm sec},&\nonumber\\
&a_3 \sim 4.0 \times 10^{-5} \ {\rm sec},&
\end{eqnarray}
where $N_{\rm chip} = 40$ is the number of the chips,
and $a_2$ and $a_3$ depend on the speed of the slave machine
(see Okumura et al. 1993; but the clock period of
the GRAPE-3AF chip is 20 MHz).
For calculating only gravity forces, the master task has to send positions
of particles (3$\times 4N$ byte) and masses (1$\times 4N$ byte),
and then the slave task has to send gravity forces (3$\times 4N$ byte) and
potential energies (1$\times 4N$ byte). Thus we estimate
\begin{equation}
a_4 = 32/r_{\rm LAN},
\end{equation}
where $r_{\rm LAN}$ is the effective transfer rate of the LAN
which depends on the size of the message.

\begin{figure}[t]
\begin{center}\leavevmode
\epsfxsize=0.8\columnwidth
\epsfbox{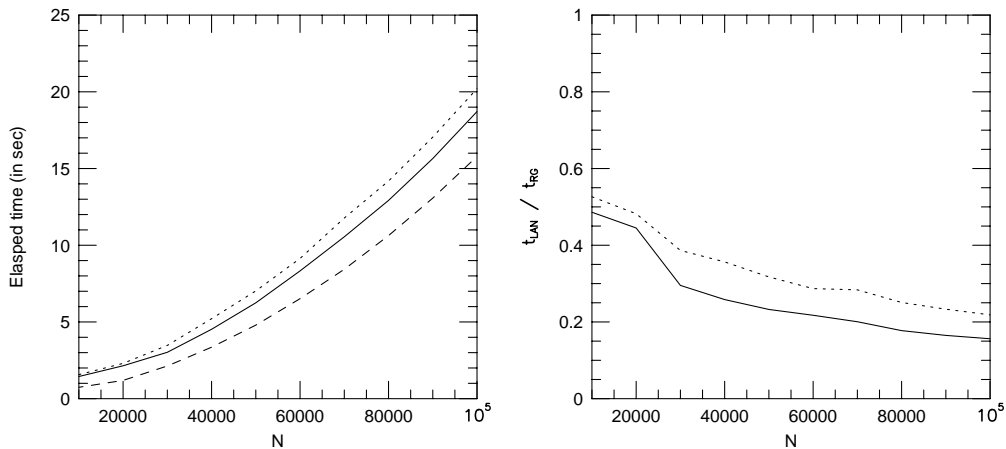}
\end{center}
\caption{
(a) left : The elapsed time as a function of $N$ in 
calculating gravitational force using Remote-GRAPE (Case RG)
for 100Mbps (solid), 10Mbps (dotted),
and usual GRAPE from the slave machine (dashed).
(b) right : The ratio between the time spent for the network
over the to total elapsed time ($t_{\rm LAN}/t_{\rm RG}$)
as a function of $N$.
\label{ratio}
}
\end{figure}

The results of our experiments are given in Figure \ref{ratio}.
Figure \ref{ratio}a shows the relation
between the elapsed time and the number of
particles, $N$, where the solid, dotted, and dashed lines represents
$t_{\rm RG}$ for 100Mbps, $t_{\rm RG}$ for 10Mbps, and $t_{\rm G}$,
respectively.
We obtain that $r_{\rm LAN}$ for this figure is
500 - 1000 Kbyte sec$^{-1}$ for 100Mbps, and 500 - 700 Kbyte sec$^{-1}$
for 10Mbps.
For 100Mbps, we can not get high performance as expected before.
This is because the effective transfer rate depends on several
factors, e.g., the state of the LAN, the speed of both machines and their bus,
and most strongly the setting and quality of the software
(operating system and device driver). 
Figure \ref{ratio}b shows the experimental relation between
$t_{\rm LAN}/t_{\rm RG}$ and $N$ for 100Mbps and 10 Mbps
by the solid and dotted, respectively.
The particles used in this experiment are the same as in Figure \ref{data}.
It is clearly seen that the ratio decreases as $N$ increases.
This means that the calculations with Remote-GRAPE
take almost the same time as with usual GRAPE when $N$ is large.
Also in this case, the performance of the slave machine does not 
affect the performance of Remote-GRAPE as shown in Section 2.

\subsubsection{Case RGN : Gravity and Neighbor}
To construct neighbor lists with Remote-GRAPE, we need additional
calculations (see Section 2) and data transfer.
In this case, the total required time are estimated as,
\begin{equation}
t_{\rm RGN} = t_{\rm GN} + t_{\rm LAN'},
\end{equation}
where
\begin{equation}
t_{\rm GN} = t_{\rm G} + [(a_5 + a_6) N_{\rm neighbor}] N,
\end{equation}
and
\begin{equation}
t_{\rm LAN'} = t_{\rm LAN} + a_7 N.
\end{equation}
Here $a_5$ is the additional time required to read neighbor lists
via VME bus, and 
$a_6$ the time spent for construction of neighbor lists for one particles;
$a_7$ is the time to transfer
neighbor lists via the LAN for one particle, which is expressed as
\begin{equation}
a_7 = a_8 \times N_{\rm neighbor},
\end{equation}
with $N_{\rm neighbor} \sim 30 - 100$
being the mean number of neighbor particles for one particle.

The original GRAPE library reads the two buffers for neighbor lists twice each.
Thus it has to transfer 4(times)$\times N_{\rm neighbor} \times$
4(one word) = 16 $N_{\rm neighbor}$ byte for one particle.
If the effective transfer rate via VME bus is 3.5 Mbyte sec$^{-1}$,
$a_5$ is estimated as
\begin{equation}
a_5 = 16/(3.5 \times 10^{6}) = 5.6 \times 10^{-5} \ {\rm sec}.
\end{equation}
It is difficult to estimate $a_6$,
because it depends highly on the integer performance of the slave machine.
However, we may estimate $[(a_5 + a_6) N_{\rm neighbor}]$ from the result of
Figure \ref{grape1} and \ref{grape2}
as $[(a_5 + a_6) N_{\rm neighbor}] \sim 5.0 \times 10^{-4}$ sec for Classic and
$\sim 2.4 \times 10^{-4}$ sec for SS-10 with $N_{\rm neighbor} \sim 40$. 
This implies that $a_5 + a_6$ depends linearly on the value of SPEC int92,
$b_{\rm SPEC}$, of the sample machines as
$(a_5 + a_6) \sim (1.9 \times 10^{-5} - 2.5 \times 10^{-7} b_{\rm SPEC})$ sec.
This leads to
\begin{equation}
a_6 \sim (1.2 \times 10^{-5} - 2.5 \times 10^{-7} b_{\rm SPEC}) \ {\rm sec}.
\end{equation}
Since we use the array of 2byte for neighbor lists,
we estimate
\begin{equation}
a_8 = 2/r_{\rm LAN} \ {\rm sec}.
\end{equation}
This leads to $a_7/a_4 = N_{\rm neighbor}/16$,
so that $t_{\rm LAN'}$ depends almost linearly on $N_{\rm neighbor}$.

\begin{figure}[t]
\begin{center}\leavevmode
\epsfxsize=0.8\columnwidth
\epsfbox{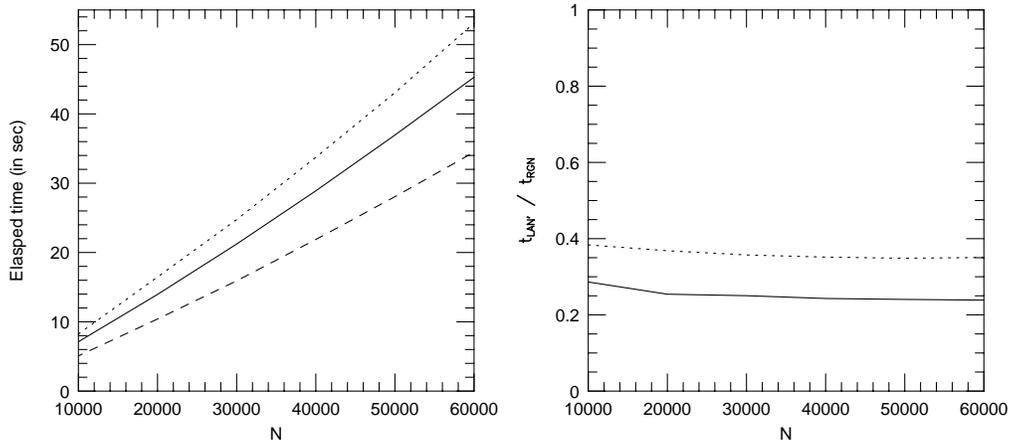}
\end{center}
\caption{
Same as Figure \ref{ratio} but for case RGN,
calculating gravitational force and neighbor lists.
\label{ratio_neighb}
}
\end{figure}

The performance of Remote-GRAPE in making neighbor lists is shown in
Figure \ref{ratio_neighb}.
Figure \ref{ratio_neighb}a shows the elapsed time as a function of $N$
as in Figure \ref{ratio}a.
Figure \ref{ratio_neighb}b shows the experimental relation between
$t_{\rm LAN'}/t_{\rm RGN}$ and $N$.
The particles used in this experiment are the same as in Figure \ref{data}.
In our configurations, the fraction of time spent for communication 
between the master and the slave is 
about 30 \% of the total time for 100 Mbps and 40\% for 10 Mbps.
In other words, the overhead is acceptable.
As discussed in section 2, however, the time required on the slave machine
(shown by the dotted line in Figure \ref{ratio_neighb}b) 
can be shorten by a factor two or more by adopting a faster slave computer.
Thus whether using Remote-GRAPE is advantageous
depends on the configuration and the nature of the simulation
(see the next section).

\begin{table*}[t]
\caption{Comparison between Remote-GRAPE and Tree method.
\label{table_tree}
}
\begin{center}
\begin{tabular}{ccccc}
\hline\hline
& Remote-GRAPE & & Tree ($\theta$)$^{\rm b}$ & \\
$\alpha$$^{\rm a}$ & & (0.8) & (1.0) & (1.5) \\
\hline
1.0  & 7.64 & 12.47 &  7.61 &  2.91 \\
5.0  & 7.63 & 19.34 & 12.79 &  6.92 \\
10.0 & 7.64 & 20.56 & 14.38 &  8.52 \\
\hline
\end{tabular}
\\
The elapsed time (sec) is shown for N = 50,000.\\
$^{\rm a}$For lager $\alpha$, clustering is larger.\\
$^{\rm b}$Three cases fo $\theta$ = 0.8, 1.0, and 1.5, 
where $\theta$ is defined in Barnes \& Hut (1986).\\
\end{center}
\end{table*}

\subsubsection{Comparison with the tree method}
Here we compare the performances of Remote-GRAPE and the tree method
(Barnes \& Hut 1986) on the master side.
We use the tree code, which is developed by J. Barnes
and freely available from the net, with a little modification.
The comparison between the two methods
on our machine is shown in Figure \ref{performance} of
the elapsed time vs. $N$.
The solid line shows the result of Remote-GRAPE and other lines
correspond to the tree methods.
The attached number indicates the parameter $\theta$
defined in Barnes \& Hut (1986).
The elapsed time of Remote-GRAPE is comparable to the tree method 
with $\theta$ = 1.0.

\begin{figure}
\begin{center}\leavevmode
\epsfxsize=0.8\columnwidth
\epsfbox{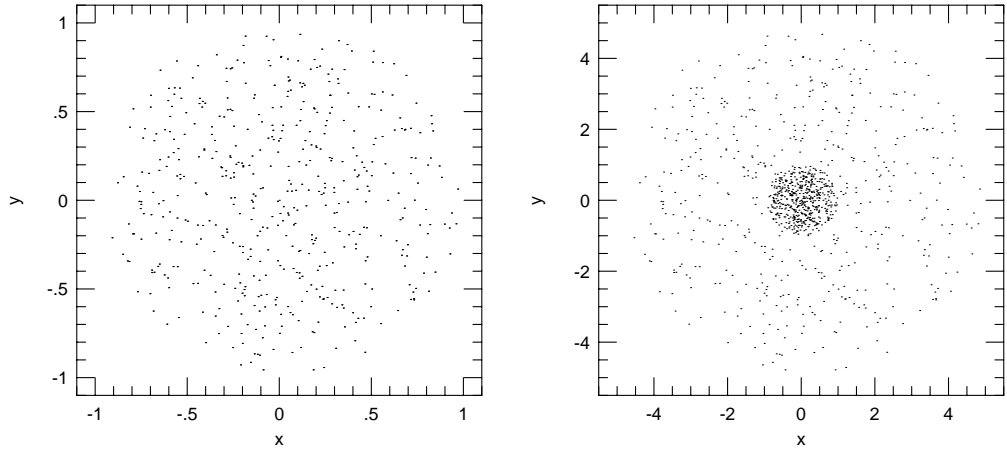}
\end{center}
\caption{
Randomly distributed points for the non-clustering state (left),
and the clustering state with $\alpha$ = 5 (right).
\label{data}
}
\end{figure}

\begin{figure}
\begin{center}\leavevmode
\epsfxsize=0.8\columnwidth
\epsfbox{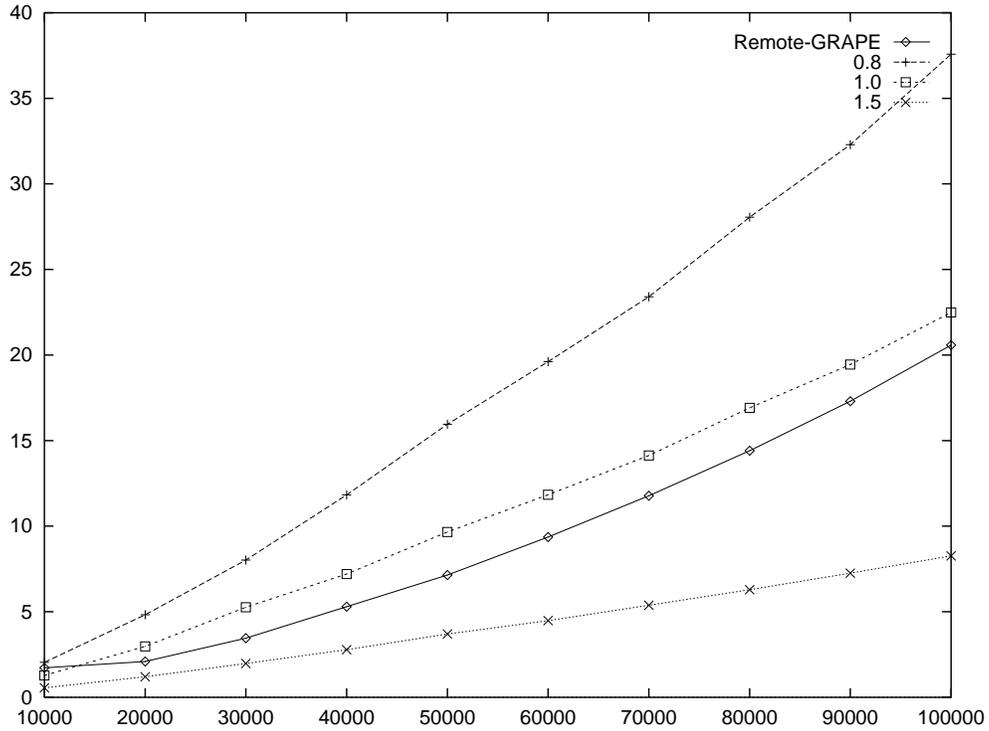}
\end{center}
\caption{
The comparison between the Remote-GRAPE and
tree code with three case of $\theta$ defined in Barnes and Hut(1986).
Shown is the elapsed time as a function of $N$.
\label{performance}
}
\end{figure}

We examine a possible advantage of Remote-GRAPE over the tree method, 
when the particles are highly clustering so that the tree is deep.
To simulate such a highly clustering situation,
we distribute particles as follows:
1) a half of the particles are randomly distributed in a certain region,
and 2) the other half are distributed in the region whose size is larger
than the former by a factor of $\alpha$ (see Figure \ref{data}).
The results for $N$ = 50,000 are shown in Table \ref{table_tree},
where $\alpha$ = 1 corresponds to the case in Figure \ref{performance}.
It is seen that for the tree method the required time
is longer for larger $\alpha$ of more strongly clustering states
because it takes longer to walk for deeper tree.
In contrast, the required time for Remote-GRAPE does not depend on $\alpha$.
For large $\alpha$, the total computational cost for Remote-GRAPE
is smaller than for the tree methods.
This is important because even if the initial state is not so
highly clustering, the state tends to get more and more strongly clustering
as the system evolves.

\section{Performance of SPH with Remote-GRAPE}

\begin{table*}[t]
\caption{
Comparison of the performance of SPH code with usual GRAPE (S-GN)
and Remote-GRAPE (S-RGN).
\label{table_sph1}
}
\begin{center}
\begin{tabular}{cccccc}
\hline\hline
Case & N$^{\rm a}$ & $t_{\rm GRAPE}$$^{\rm b}$ sec (\%) &
$t_{\rm SPH}$$^{\rm b}$ sec (\%) &
$t_{\rm misc}$$^{\rm b}$ sec (\%) &
Total$^{\rm b}$ sec \\
\hline
S-GN  & 5000  &  2.1 (14) & 12.1 (84) & 0.3  (2) & 14.5 \\
S-RGN & 5000  &  3.3 (83) &  0.6 (17) & 0.02 (0) &  3.9 \\
S-GN  & 10000 &  4.2 (15) & 23.1 (83) & 0.6  (2) & 27.9 \\
S-RGN & 10000 &  6.2 (83) &  1.2 (16) & 0.04 (1) &  7.4 \\
\hline
\end{tabular}
\\
$^{\rm a}$ $N$ is the number of gas particles.\\
$^{\rm b}$ $t_{\rm GRAPE}$, $t_{\rm SPH}$, and $t_{\rm misc}$ are
the elapsed time in gavity plus neighbor, sph, 
and miscellaneous part, respectively, and their percentage are given
in parenthesis. Total is the total elapsed time for one step.\\
\end{center}
\end{table*}

\subsection{General performance}
\begin{figure}[t]
\begin{center}\leavevmode
\epsfxsize=0.8\columnwidth
\epsfbox{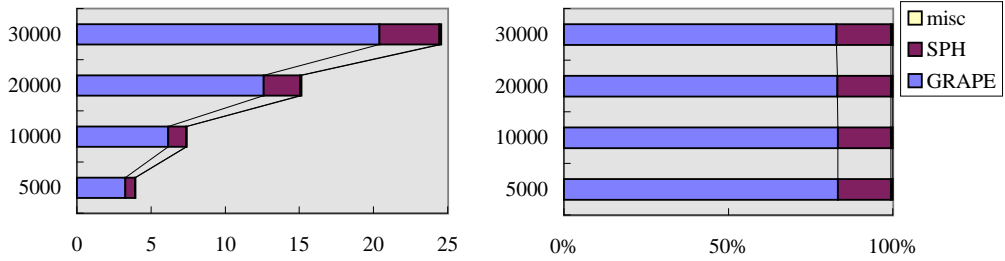}
\end{center}
\caption{
The performance of our SPH code using Remote-GRAPE (case S-RGN)
for different $N$. (a) left : The elapsed time.
(b) right : The fraction occupied by different part of the code.
\label{sph}
}
\end{figure}

In order to see the actual advantage of SPH with Remote-GRAPE
(hereafter Case S-RGN), 
we make a comparison with the performance of SPH that uses
usual GRAPE on the slave machine (Case S-GN). 
The detail description of our SPH code is presented in Mori et al. 1996.
In Table \ref{table_sph1}, we show the computing time spent in various
parts of the calculation per step, namely,
gravity and neighbor ($t_{\rm GRAPE}$),
SPH ($t_{\rm SPH}$), and misc ($t_{\rm misc}$),
where $t_{\rm GRAPE}$ is the time to calculate gravity and
neighbor lists, and $t_{\rm SPH}$ and $t_{\rm misc}$ are, respectively,
those for SPH calculations and miscellaneous calculations.
The time spent on the host computers (
$t_{\rm SPH+misc} = t_{\rm SPH} + t_{\rm misc}$),
which has been the bottleneck of the performance in previous GRAPE-SPH
(Steinmetz 1996), differs 20 times between two cases.
In Case S-RGN, we compare $t_{\rm GRAPE}$, $t_{\rm SPH}$, and $t_{\rm misc}$
for larger $N$ in Figures \ref{sph}a and \ref{sph}b.
For the adopted configuration, the time spent for gravity and neighbor part
occupies the largest fraction of the total time (i.e., $\sim$ 80 \%), and 
the fractions of the three parts are almost independent of $N$.
The large fraction of $t_{\rm GRAPE}$ make the total time per step 
be $\sim$ 4 times shorter in Case S-RGN than Case S-GN
despite the use of the 20 times faster host computer.
This implies that in Case S-RGN, the bottle-neck of the performance is
now the $t_{\rm GRAPE}$ part rather than the $t_{\rm SPH}$ part.

We summarize the results of Section 3.3.1 \& 3.3.2 to
present the $N$ dependence of $t_{\rm GRAPE}$ in Case S-RGN as follows :
\begin{eqnarray}
&t_{\rm GRAPE} = t_{\rm RGN} = t_{\rm GN} + t_{\rm LAN'}&\nonumber\\
&= a_1 N^2 + [a_2 + a_3 + a_4 + (a_5 + a_6 + a_8) N_{\rm neighbor}]
N &\nonumber\\
&= 5.0 \times 10^{-8} N^2/N_{\rm chip} + [6.0 \times 10^{-5} + &\nonumber\\
&(32+2 N_{\rm neighbor})/r_{\rm LAN} + &\nonumber\\
&(16/r_{\rm BUS} + (1.2 \times 10^{-5} - 2.5 \times 10^{-7}
b_{\rm SPEC})) &\nonumber\\
&N_{\rm neighbor}] N \ \rm{sec},&
\end{eqnarray}
where $r_{\rm BUS}$ is the speed of the bus in byte per second.
For $a_3$, we use the value of Okumura et al.(1993),
though its machine dependence is not known.
If we adopt a typical value for our configuration, 
we get
\begin{equation}
t_{\rm GRAPE} = 1.3 \times 10^{-9} N^2 + 7.5 \times 10^{-4} N \ {\rm sec},
\end{equation}
which is consistent with the actual result of $t_{\rm GRAPE}$.
The additional time for the calculation of neighbor lists is,
\begin{eqnarray}
&t_{\rm neighbor} = [((a_5 + a_6) + a_8) N_{\rm neighbor}] N  =&\nonumber\\
&= [(5.0 \times 10^{-4}) + (1.4 \times 10^{-4})] N &\nonumber\\
&= 6.4 \times 10^{-4} N \ {\rm sec}.&
\end{eqnarray}
Thus we can estimate $t_{\rm neighbor}/t_{\rm GRAPE} \sim 0.8$
for $N \sim 10000 - 60000$.
Possible ways to overcome the bottleneck would be:
(1) calculating neighbor lists with a faster slave machine,
(2) using Remote-GRAPE from time to time to get neighbor lists
by adopting a larger neighbor radius, 
and (3) using another method to construct neighbor lists on the master side.

\subsection{Parallel method}
The performance of GRAPE-SPH can be further improved
if we compute gravity and neighbor lists,
and other parts in parallel by using Remote-GRAPE.
We have implemented a parallel version of routines,
where the sample code is as follows:
\begin{verbatim}
.........
call g3setscales(scale_leng, scale_mass)
call g3seteps2(eps2)
call g3setmass(nall,mass)
call g3setdata(nall,x(1,1), x(2,1), x(3,1))
call g3seth2all(nall, h)
call g3docalculationpara(1,nall,1, nall, 2)
.........
another calculations
.........
call g3getgravityasync(nall, acc(1,1), acc(2,1),
     acc(3,1), pot)
call g3getneighborasync(nall, list, begin)
\end{verbatim}
Three routines are replaced with parallel routines.
Firstly, all required data in the slave side are sent
by the same routines as given in Section 3.2.
The difference between {\tt g3docalculation} and {\tt g3docalculationpara}
is when it returns. {\tt g3docalculation} returns after all the calculation
on the slave side is done, while, {\tt g3docalculationpara} returns
after the slave task starts the calculation.
The master side starts other calculations after
{\tt g3docalculationpara} returns.
The two routines, which are denoted by ``async'' attached to those names,
receive the message asynchronously.
When we call {\tt g3getgravityasync} and if the message,
which contains gravity, has not arrived,
it waits for the arrival of the message.
{\tt g3getneighborasync} works in the same way.

As mentioned in Section 2, we need neighbor lists to compute
physical quantities in SPH.
In applying the parallel routines for the SPH calculation, 
we compare several methods to make neighbor lists.
We summarize the test cases for GRAPE-SPH in Table \ref{table_sph2}.

\begin{table*}[t]
\caption{Description of test cases for GRAPE-SPH.
\label{table_sph2}
}
\begin{center}
\begin{tabular}{cccccc}
\hline\hline
Case & Gravity & Neighbor lists &
SPH part & Parallel$^{\rm a}$ & Section$^{\rm b}$ \\
\hline
S-GN   & slave & slave  & slave  & $\times$ & 4.1  \\
S-RGN  & slave & slave  & master & $\times$ & 4.1  \\
PS-RG  & slave & master & master & $\bigcirc$  & 4.2.1\\
PS-RGN & slave & slave  & master & $\bigcirc$  & 4.2.2\\
\hline
\end{tabular}
\\
''slave'' and ``master'' in each column represent respectively
that the part is done on ``slave'' or ``master'' machine.\\
$^{\rm a}$This column indicates whether parallel or not.\\
$^{\rm b}$This column represents which section the test case is discussed.\\
\end{center}
\end{table*}

\subsubsection{Case PS-RG : Parallel SPH using Remote-GRAPE for gravity only}
\begin{figure}[t]
\begin{center}\leavevmode
\epsfxsize=0.8\columnwidth
\epsfbox{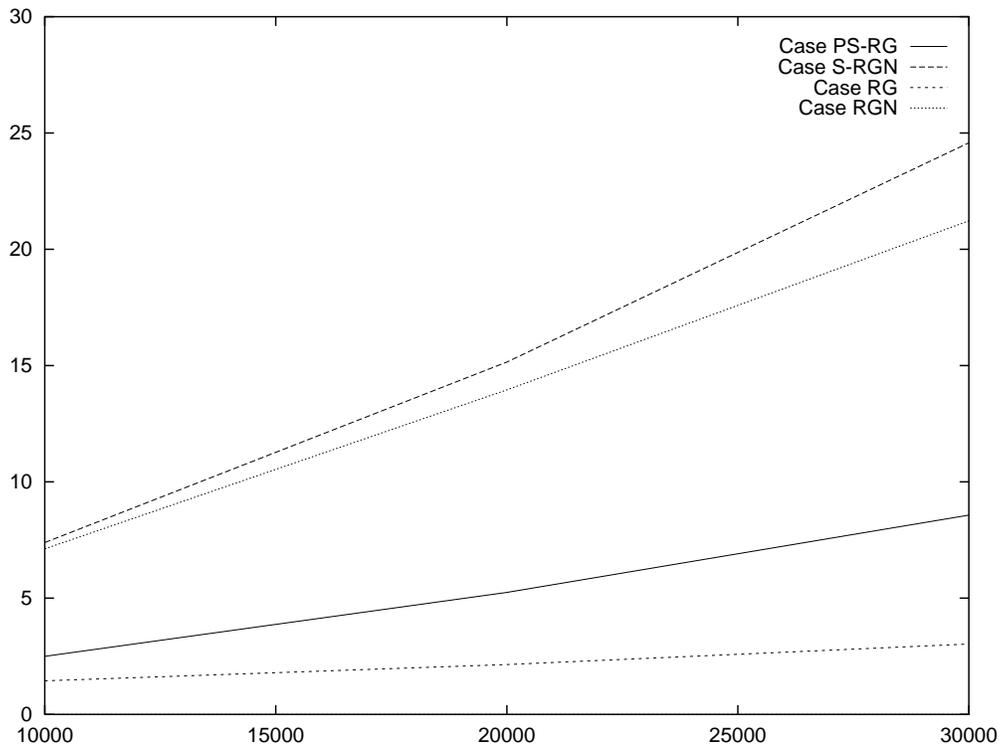}
\end{center}
\caption{
The elapse time per step as a function of $N$ in
Case PS-RG (solid) and S-RGN (dashed).
The dotted and short-dashed lines correspond respectively to
case RG and case RGN.
\label{para}
}
\end{figure}

As shown in the previous section, it is more advantageous
to use Remote-GRAPE for gravity only.
We first examine the method to construct neighbor lists on the master side
(Case PS-RG). Results are shown in Figure \ref{para},
where the solid and dashed lines correspond to the Case PS-RG
and Case S-RGN, respectively.
Other two lines show Case RG (dotted) and Case RGN (short-dashed)
for comparison.
The enhancement of the performance is rather dramatically.
In Table \ref{table_sph3}, we present comparison of Case S-GN, S-RGN,
and PS-RG for $N \sim 10000$.
The performance of case PS-RG is 3 times faster than case S-RGN, 
10 times than case S-GN.
However, we use the tree structure to construct neighbor lists,
so that the clustering state apparently affect the performance
because of the same reason mentioned in Section 3.3.3.
It takes 2 sec to construct neighbor lists on the master side
if $\alpha \sim 1$ ($N \sim 10000$).
For the deeper tree, it takes longer time ($\sim$ 20 sec).

\begin{table*}[t]
\caption{Comparison of the performance of Case S-GN,
 Case S-RGN and Case PS-RG.
\label{table_sph3}
}
\begin{center}
\begin{tabular}{cc}
\hline\hline
Case & Total sec\\
\hline
S-GN  & 27.9 \\
S-RGN & 7.4  \\
PS-RG & 2.5  \\
\hline
\end{tabular}
\\
The elapse time for one step is shown. $N \sim 10000$.
\end{center}
\end{table*}

\subsubsection{Case PS-RGN : Using Remote-GRAPE for gravity and neighbor lists}
Another possible method is to use the neighbor lists of the previous step,
which is constructed by Remote-GRAPE,
as possible neighbor lists (Case PS-RGN).
In this case, we have to adopt larger neighbor radius than Case S-RGN.
Figure \ref{sph}b indicates that the $t_{\rm GRAPE}$ part of the calculation
is dominant so that the parallel methods has less advantage
for our configuration.
Since the $t_{\rm GRAPE}$ part and the other parts of the calculation
obey different $N$ dependence, it is worth examine if 
there exits an optimal number of particles that leads to 
the optimal performance for a given configuration in Case PS-RGN.
For the SPH and misc part, we estimate their $N$ dependence
in our SPH code from Figure \ref{sph}a.
We get $t_{\rm SPH+misc} = 1.8 \times 10^{-4} N$ sec.
Therefore, $t_{\rm GRAPE}$ always exceeds $t_{\rm SPH+misc}$ for any $N$.

However, Figure 9 and Table \ref{table_sph1} are the results for
pure SPH calculations.
Because we are interested in the formation of galaxy and globular cluster,
we include such physical processes in our SPH code as cooling, heating,
and star formation (Mori et al. 1996).
If we include these physical processes, $t_{\rm misc}$ should be much longer,
so that the parallel method has more advantage.
For example, in the SPH code calculation with
non-equilibrium H$_2$ cooling, it takes $\sim 2.7 \times 10^{-4}$ sec
to solve rate equations for 10 species on our machine.
In many cases, the cooling time scale is shorter than the dynamical time scale,
so that the rate equations need to be solved many times in one dynamical step,
which leads to $t_{\rm misc} > 10^{-3} N$ sec in our machine.
For large $N$, $t_{\rm misc}$ is dominant than $t_{\rm GRAPE}$.

\section{Conclusion}
We describe the Remote-GRAPE library and analyze its performance.
It allows us to use GRAPE-3A with the computer
which is not directly connected to GRAPE.
Thus, we can use the state-of-the-art computer as a host computer.

Firstly, we analyze the performance of original GRAPE system in some detail
as summarized below:
\begin{enumerate}
\item
If we calculate only gravity forces, the use of larger number
of boards leads to higher performance.

\item
In calculating both gravity and neighbor lists, however,
the computation required on host WS is larger than the former case, 
so that the increase of number of the boards is less advantageous.
 
\item
The speed of the host WS dramatically change the performance of GRAPE system.
\end{enumerate}

Secondly, we analyze the performance of Remote-GRAPE for our configuration
as summarized below : 
\begin{enumerate}
\item
If we calculate only gravity forces,
the time required by Remote-GRAPE is almost the same as that of original GRAPE.

\item
In calculating both gravity and neighbor lists by Remote-GRAPE,
the time required to transferring data does not occupy
a large fraction of total time but $\sim$ 20-30 \% for our configuration.

\item
Compared with the tree method, the advantage of Remote-GRAPE is that
its performance does not depend on the state of clustering, 
while the required time of the tree method is longer
for higher degree of clustering.

\item
In the actual application, the performance of our SPH code using Remote-GRAPE
is 3 - 4 times faster than the SPH code using original GRAPE 
on the slave machine.

\item
We can get further high (10 times than the SPH on slave machine)
performance by using the parallel method with Remote-GRAPE
to calculating only gravity. 
\end{enumerate}

We have completed the first version of Remote-GRAPE, 
which significantly improves the performance of GRAPE-SPH. 
However, there are several ways to improve further the performance of our
library and GRAPE-SPH with Remote-GRAPE.
\begin{enumerate}
\item
If we use a larger number of GRAPE boards with a fast bus and 
a fast slave machine, the use of GRAPE in constructing
neighbor lists is more advantageous than the use of tree structure.

\item
We can develop more efficient library.
In the GRAPE system all values are converted to the fixed floating point
format in the original GRAPE library and then sent to the GRAPE board.
The conversion from the floating point format to the fixed format
may be done by the master side. In a relatively slow slave computer,
this will significantly improve the performance.
\end{enumerate}

We would like to thank J. Makino and T. Suzuki for reading the manuscript and
giving useful comments.  
We also thank T. Miwa for providing us the result with SS-10.
Those who would like to use our library can contact authors
(nakasato@astron.s.u-tokyo.ac.jp).
This work has been supported in part by the grant-in-Aid
for Scientific Research (05242102, 06233101) and COE research (07CE200)
of Ministry of Education, Science, and Culture of Japan.

\end{document}